\documentclass[a4paper, oneside, twocolumn, notitlepage, 10pt]{extarticle_ecoc}
\usepackage{ecoc}
\usepackage{caption}
\usepackage{subcaption}
\begin{document}
\selectlanguage{english}    

\title{Spectral Power Profile Optimization of Field-Deployed WDM Network by Remote Link Modeling}%

\author{
    Rasmus T. Jones\textsuperscript{(1)},
    Kyle R. H. Bottrill\textsuperscript{(2)},
    Natsupa Taengnoi\textsuperscript{(2)},
    Periklis Petropoulos\textsuperscript{(2)},
    Metodi P. Yankov\textsuperscript{(1)}
}

\maketitle

\begin{strip}
 \begin{author_descr}

   \textsuperscript{(1)} Department of Photonics Engineering, Technical University of Denmark,
   \textcolor{blue}{\uline{rajo@fotonik.dtu.dk}}

   \textsuperscript{(2)} Optoelectronics Research Centre, University of Southampton,
   \textcolor{blue}{\uline{k.bottrill@soton.ac.uk}}

 \end{author_descr}
\end{strip}

\setstretch{1.1}

\renewcommand\footnotemark{}
\renewcommand\footnoterule{}

\begin{strip}
  \begin{ecoc_abstract}
    A digital twin model of a multi-node WDM network is obtained from a single access point. The model is used to predict and optimize the transmit power profile for each link in the network and up to 2.2~dB of margin improvements are obtained w.r.t. unoptimized transmission.  \textcopyright2022 The Author(s)
  \end{ecoc_abstract}
\end{strip}

\section{Introduction}
Wide-band wavelength division multiplexing (WDM) systems and all optical amplification are cornerstones of our modern digital information infrastructure and are responsible for today's transmission data rates.

The nonlinear wavelength dependence of the gain and noise on each channel in a WDM system is a key detrimental effect governing the quality of transmission per user in WDM networks\cite{perin2019importance}. In wide-band WDM systems, wavelength dependent effects such as transceiver penalties, nonlinear fiber effects, and non-flat erbium-doped fiber amplifier (EDFA) gain and noise contributions are non-trivial to model\cite{saleh1990modeling}.

Channel effects are reasonably well captured by analytical models \cite{poggiolini2012gn}.
However, accurate representation of EDFAs using analytical models is challenging.
Machine learning (ML)\cite{morette2021leveraging,morette2022robustness,koch2022reinforcement} and data aided EDFA models \cite{yankov2021power,yankov2021snr,kamel2021osnr,meseguer2021highly,sena2022link,jiang2022machine} are alternatives that gather attention from the community. Training an ML model for an EDFA and other link components requires full access to its input and output in a lab environment. Yet, deployed fiber links and their respective components cannot be physically isolated and used for data driven modeling approaches. Further, aging effects of an EDFA device are difficult to measure and model after the device has been deployed.

We present a novel way of training an EDFA ML model remotely, where the EDFA device is deployed in a network of links. Under the assumption that all EDFA devices in the field-deployed network are of the same manufacturer and make, we only train on a single EDFA device and use it to model all other EDFA devices in the network.
We thereby demonstrate that an entire, deployed optical fiber network and its individual components can be modeled by only sending probe signals through a single access point on a single point to point link.
The model is then used to optimize operation specific properties of the network\cite{roberts2016convex,ionescu2022optimization, garbhapu2021network}. In particular, the spectral power per WDM channel is optimized with a target of equalizing the signal to noise ratio (SNR) across all channels without requiring feedback on the quality of transmission as in Garbhapu et al.\cite{garbhapu2021network} and without requiring gain flattening filters after the EDFAs\cite{yankov2021snr}.

\begin{figure}[t]
   \centering
        \includegraphics[width=0.9\linewidth]{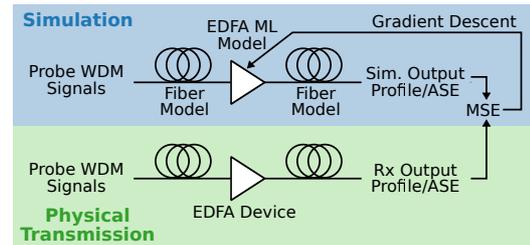}
    \caption{Remote modeling of EDFA device.}
    \label{fig:training}
\end{figure}

\section{Remote EDFA Modeling and Optimization}
The methods proposed extend the methodology in our previous work\cite{yankov2021power,yankov2021snr} performed in a lab environment. The overall goal is to predict the signal power and the accumulated noise power for all WDM channels of a link as a function of its input power profile. The model is then used to optimize for operation-specific properties of the link via gradient descent.

A link model is comprised of analytical fiber models and trained EDFA ML models in cascade, appended with a differentiable interpolation model for the wavelength dependent implementation penalties of the transmitter-receiver (TRX) chain. The EDFA model predicts the amplifier gain profile and the profile of the amplified spontaneous emission (ASE) added by the amplifier as a function of the input power profile and the amplifier operating point (its total input and total output power).

In contrast to our previous work\cite{yankov2021snr}, where the EDFA is entirely modeled by neural networks (NNs), in this work a combination of a physical model and NNs is used. More specifically, the physical EDFA model presented in Meseguer et al.\cite{meseguer2021highly} is modified, replacing both look up tables for the total power dependent gain profile and noise figure functions with NNs, ensuring that the resulting EDFA model is differentiable. Model training is performed in the following way, exemplified in Fig.~\ref{fig:training}. The EDFA model weights are initialized, and a cascade digital twin model of the fiber-EDFA-fiber link is created. Probe signals are sent on the link, and the system response is measured. The input probe signals consist of a shaped ASE spectrum with a random power profile. The spectrum is shaped so as to emulate a C-band, 48-channel, 12.5~GBd WDM signal on a 100 GHz grid. The response measurements consist of wavelength dependent power profile and ASE noise added at each channel of interest, measured using an optical spectrum analyzer (OSA). The mean squared error (MSE) between the model prediction and the measurements is estimated and used as cost for gradient descent-based EDFA weight updates. 
The analytical part of the EDFA hybrid model ensures that the EDFA weights are only responsible for modeling EDFA effects. In a way, the NNs in the EDFA ML model are naturally regularized against overfitting to link-specific effects.

The obtained EDFA ML model is thereafter used in cascade with fiber models and a TRX penalty model for performance prediction and gradient descent optimization of the input power profile to a chosen link of the network. The chosen cost function is  $Cost~=~-\text{min}_{\lambda}(\text{SNR}(\lambda))$, targeting a flat SNR.

\section{Experimental Setup}
\begin{figure}[t]
   \centering
        \includegraphics[width=\linewidth]{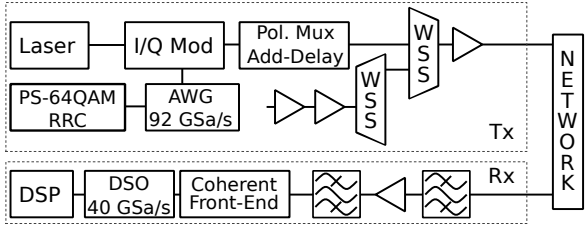}
    \caption{Experimental transceiver setup at network node Southampton, STN.}
    \label{fig:txrx}
\end{figure}
\begin{figure}[t]
   \centering
        \includegraphics[width=\linewidth]{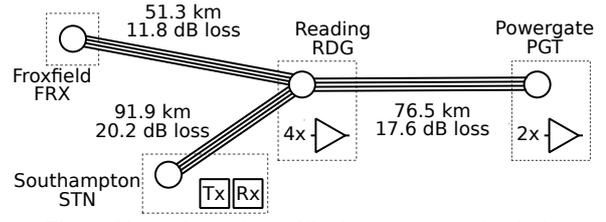}
    \caption{Network topology. All edges are comprised of 4 individual fiber links.}
    \label{fig:network}
\end{figure}

The experimental setup is comprised of a transceiver and a network of fiber links, as depicted in Fig. \ref{fig:txrx} and Fig. \ref{fig:network}, respectively. The transmitter emulates a 48 channel WDM system with a channel spacing of 100~GHz. A setup of EDFAs and WSSs emulate 47 WDM channels from an ASE noise source of bandwidth 12.5~GHz, and a single data channel is loaded with a modulated signal, which is multiplexed into the system on the wavelength under test. The data channel uses 64QAM with probabilistic amplitude shaping (PAS) \cite{bocherer2015}, low-density parity check forward error correction (FEC) with overhead of 33$\text{\%}$, root-raised-cosine pulse shaping with 0.3 roll-off at a baud rate of 15.33~GBd. Polarization multiplexing is emulated with a standard delay-and-add technique. The entropy of the constellation is chosen to 5.5 bits/symbol, which is near-optimal for the range of expected link SNRs between 14 and 18~dB. The WSSs are further used to apply an arbitrary power profile across the WDM channels which are thereafter launched into the network at a total power of 18~dBm. The latter was found optimal for the scenarios under consideration. The network is composed of 4 nodes at Southhampton (STN), Reading (RDG), Froxfield (FRX) and Powergate (PGT), with 6 EDFAs in total - 4 in RDG and 2 in PGT. The accessing node is located at STN. At the receiver, the data channels are extracted with two optical filters and an EDFA, followed by a coherent front-end and a digital storage oscilloscope (DSO). The DSP is pilot-based as reported in our previous work\cite{yankov2021snr}.

For the experimental demonstration, two links of the network are chosen. First, a link of 439.4~km length, with a path configuration as follows: STN-RDG*-PGT*-RDG*-FRX-RDG*-STN (a * denotes an amplification node). Second, a longer link of 592.4~km length is chosen, with a path configuration as follows: STN-RDG*-PGT*-RDG*-PGT*-RDG*-FRX-RDG*-STN, where the duplicating hops between RDG and PGT use independent fiber links. Overall, in the first and second link the signal is amplified 4 and 6 times, respectively, by independent EDFAs. The EDFAs are model CEFA-644-00 from Lea Photonics. The fiber links are composed of standard, single mode fiber with distances and total losses given in Fig.~\ref{fig:network}

\begin{figure*}[t]
    \centering
    \begin{subfigure}[t]{0.49\textwidth}
        \centering
        \includegraphics[width=\textwidth]{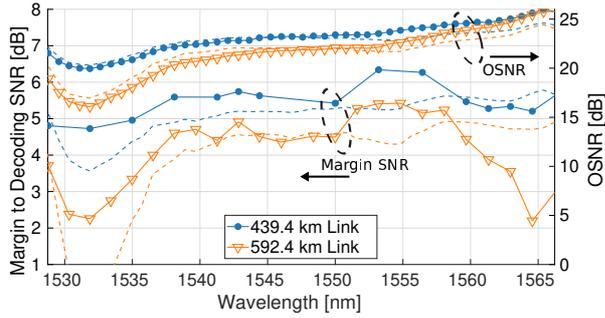}
    \end{subfigure}
    \hfill
    \begin{subfigure}[t]{0.455\textwidth}
        \centering
        \includegraphics[width=\textwidth]{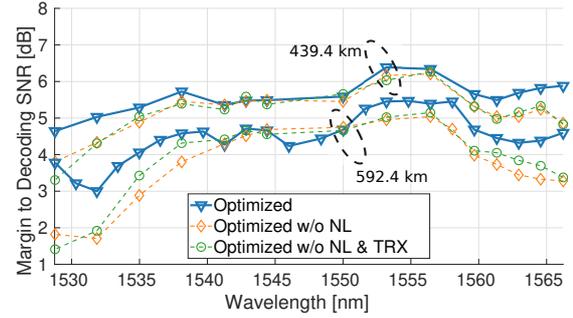}
    \end{subfigure}
    \caption{\textbf{(left)} SNR margin to the SNR decoding threshold and OSNR measurements (solid lines) and predictions (dashed lines) for a flat input power profile for the long and short link. \textbf{(right)} SNR margin of the optimized profiles for the long and short link for 3 performed profile optimizations, including all impairments (solid), excluding NL (dashed), excluding NL and TRX (dashed).}
    \label{fig:results}
\end{figure*}

For the corresponding digital twin model, only a single EDFA device at the RDG node is characterized with the method described in the previous section, through a link with a path configuration as follows: STN-RDG*-STN (independent fibers in both directions). The obtained EDFA ML model is used to predict the response of all 6 EDFAs in the network. The model for the wavelength-dependent TRX SNR impairments is obtained from back-to-back measurements of the system.
The digital twin model, including fiber models, EDFA ML models and TRX model, is used to optimize the input power profile launched into the link, targeting a flat output SNR as described in the previous section.
Optimization is performed for digital twin models of varying levels of complexity: 1) including all impairments; 2)~excluding stimulated Raman scattering and Kerr fiber nonlinearities (NL); 3) excluding NL and the TRX penalties. Including a flat input profile, this leads to 4 different candidate power profiles. 

\section{Results}
For two link scenarios of the presented network, the received SNR of up to 24 out of the 48 WDM channels is measured. The SNR decoding threshold for the chosen FEC and modulation format is 12.5~dB. The SNR margin to that number is of interest. In Fig.~\ref{fig:results} (left), the SNR margins and the optical SNR (OSNR) measurements (solid lines) and predictions (dashed lines) for a flat input power profile are reported for the two systems. A maximum OSNR error of $\approx$~2~dB at the long wavelength region is reported, even though the prediction model has only seen one of the EDFAs comprising the link - the one at RDG. The SNR prediction is slightly worse with up to 2.5~dB of maximum error for the long link. We attribute this additional error to the poorer performance of the receiver EDFA, which may not be ideally modeled at low input powers and the edge of the spectrum. The \textit{minimum} margin with a flat input profile is $\approx$~2.2~dB and $\approx$~4.8~dB at 1532.5~nm for the long and short link, respectively.

In Fig.~\ref{fig:results} (right), the margins of the optimized profiles are shown for the long and the short link for the 3 performed profile optimizations.
By excluding either of the NL or the TRX penalties, the margin is worse than when using a flat input profile. When the system is optimized on the digital twin model including all impairments, the \textit{minimum} margins (around 1532.5~nm) are improved by $\approx$~0.8~dB to $\approx$~3~dB and $<$~0.1~dB to $\approx$~4.8~dB, for the long and short link, respectively. In addition to the minimum margin gain, we also note the improvement of the margin in the long wavelength region by $\approx$~2.2~dB and $\approx$~0.8~dB for the long and short link, respectively.  
For comparison, the method from Garbhapu et al.\cite{garbhapu2021network} achieved 0.5~dB margin improvement over a lab setup, where performance feedback from the receiver drives the power allocation. In this work, the performance improvement is available immediately and without requiring any feedback from the network. We expect that for longer link configurations with more amplification stages, the unoptimized profile will perform even worse as suggested by the higher gains obtained for the long link than for the short link under test in this paper.

\section{Conclusions}
The proposed method allows for full modeling of field-deployed links without requiring access to all nodes nor feedback from the network on the current performance. This enables modeling and optimization of operational systems, only requiring characterization of an isolated link. The proposed optimization method achieved up to 2.2 dB of margin improvements for links of up to 6 amplified spans and 592.4 km.

\begin{footnotesize}
\section{Acknowledgements}
These experiments were carried out on EPSRC's National Dark Fibre Facility. The work received financial support by the UK's EPSRC through research projects EP/S028854/1 and EP/S002871/1, and Innovation Fund Denmark project RANON, ref. 1047-00013.
\end{footnotesize}

\newpage

\printbibliography

\vspace{-4mm}

\end{document}